# MOCVD growth mechanisms of ZnO nanorods

G Perillat-Merceroz[1,2], P H Jouneau[2], G Feuillet[1], R Thierry[1], M Rosina[1], and P Ferret[1]

[1]CEA - LETI, Minatec, 17 rue des Martyrs, 38054 Grenoble cedex 9, France
[2]CEA - INAC, 17 rue des Martyrs, 38054 Grenoble cedex 9, France

guillaume.perillat-merceroz@cea.fr

**Abstract**. ZnO is a promising material for the fabrication of light emitting devices. One approach to achieve this goal is to use ZnO nanorods because of their expected high crystalline and optical quality. Catalyst free growth of nanorods by metalorganic chemical vapour deposition (MOCVD) was carried out on (0001) sapphire substrates. Arrays of well-aligned, vertical nanorods were obtained with uniform lengths and diameters. A thin wetting layer in epitaxy with the sapphire substrate is formed first, followed by pyramids and nanorods. The nucleation of nanorods occurs either directly at the interface, or later on top of some of the pyramids, suggesting various nucleation mechanisms. It is shown that crystal polarity plays a critical role in the growth mechanism with nanorods of Zn polarity and their surrounding pyramids with O polarity. A growth mechanism is proposed to explain that most threading dislocations lie in the wetting layer, with only a few in the pyramids and none in the nanorods.

## 1. Introduction

ZnO has been studied for decades, but recently generated renewed interest due to its potential applications for optoelectronic devices [1]. Its large band gap (around 3.4 eV, similar to GaN) makes ZnO a candidate for short wave length light emitting diodes (LEDs) and laser diodes. The large exciton binding energy (60 meV compared with 25 meV for GaN) is promising for light emission at room temperature, presuming that the difficult problem of p type doping is solved. Growth of ZnO nanowires or nanorods is widely reported because of possible uses as field emitters, solar cells, gas sensors or LEDs. Their high crystalline quality and purity, due to growth without a catalyst, are adequate for optoelectronic applications. MOCVD growth [2] enables fast, large area deposition at an industrial scale. Understanding of the growth mechanism of nanorods is needed for better control of their density and size and should give an indication as to why the nanorods are defect-free.

## 2. Experimental details

ZnO nanorods were grown using catalyst-free MOCVD in a horizontal hot-wall Epigress reactor. The c-plane sapphire substrates were annealed at 1200°C in oxygen. Diethylzinc (DEZn) was used as Zn precursor, $N_2O$ as the oxidiser, with a VI/II ratio between 100 and 1000. The ZnO nanorods were grown at 750-850°C, with different deposition times in order to follow the formation of the nanorods. The morphology of the samples was investigated by scanning electron microscopy (SEM) using a Zeiss Ultra 55 microscope. Structural studies were carried out by transmission electron microscopy (TEM) on two microscopes: a 300 kV FEI Titan and a 400 kV Jeol 4000 EX.

## 3. Results and discussion

3.1. Morphology

After a growth of 20 seconds to 1 minute, a thin 2 dimensional ZnO wetting layer is observed on the sapphire substrate (as shown by cross sectional TEM in Figure 5b), together with nanorods and pyramids (Figure 1a and b). The higher density seen in the 20 seconds growth sample is due to different growth conditions. After one hour of deposition, nanorods have grown in diameter and length. At this stage, the nanorods have a typical length of 4 μm, and a diameter of 200 nm (Figure 1c). Nanorods grow along the c direction as demonstrated by X-ray diffraction [3]. Some pyramids exhibit nanorods on their top and some do not.

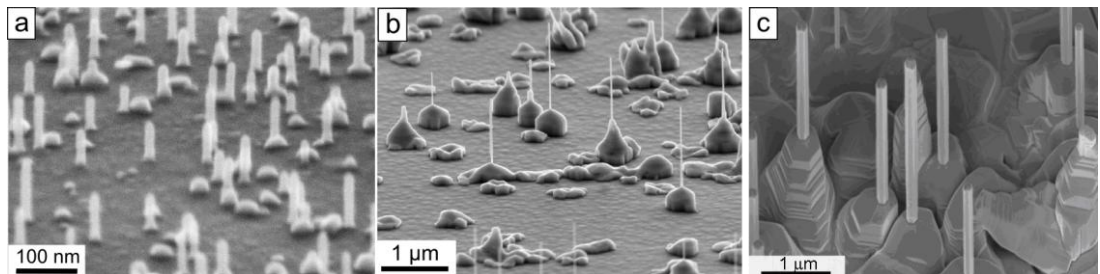

Figure 1. Tilted SEM images of ZnO nanorods: (a) after 20 s of growth, (b) after 1 min of growth, (c) after 1 hour of growth.

3.2. Role of crystal polarity

We used convergent beam electron diffraction (CBED) to determine the crystal polarity of nanorods and pyramids. Nanorods were found to be of Zn polarity while pyramids have O polarity (Figure 2). This is in agreement with a recent result by Sun et al. [4] who reported that nanorods grown by pulsed laser deposition exhibit Zn polarity. Contrary to that the buffer layer shows O polarity.

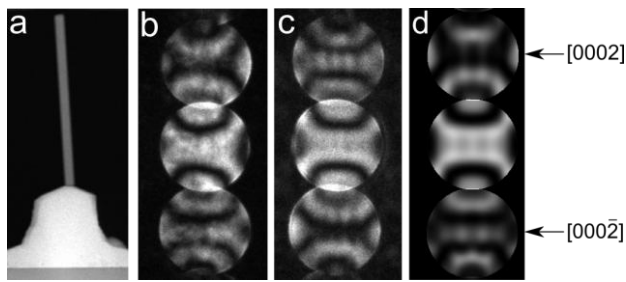

Figure 2. Determination of the polarity of both the nanorod and the pyramid: (a) STEM image of the analyzed zone, (b) and (c) respective CBED patterns of the nanorod and the pyramid. Growth direction upwards. (d) Simulated CBED pattern with the JEMS software.

Facets of the hexagonal nanorods have been indexed basing on intensity profiles of nanorods viewed along the <1-100> or <11-20> zone axis. It was found that they are of {1-100} type (Figure 3). Laudise et al. have also reported the higher growth rate of the [0001] direction (Zn polarity) and the presence of {1-100} facets in hydrothermal growth [5]. Growth of nanorods requires Zn polarity nuclei which have a high growth rate in the [0001] direction, partly because of the low free energy of the {1-100} facets.

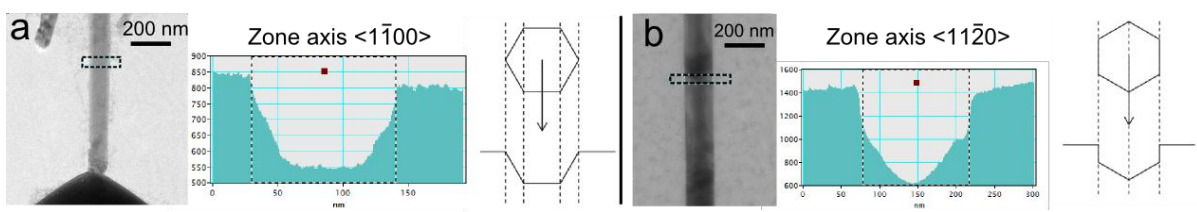

Figure 3. Identification of the side facets of the nanorods by intensity profiling of TEM bright-field images: (a) <1-100> zone axis (b) <11-20> zone axis.

3.3. Nanorod nucleation

Nanorods originate either from the ZnO / sapphire interface (Figure 4a), or from the top of some pyramids (Figure 4b). Contrasts in the nanorod are bend contours resulting from the glue-induced strain. No defects are visible in nanorods dispersed on a carbon film (Figure 5d). For one sample, pyramids and nanorods were observed to be aligned in a <11-20> direction (Figure 4c). In this case, nucleation of the pyramids and of the nanorods would be linked to atomic steps at the substrate surface [6]. However, as the alignment along a particular direction is not always seen and as some nanorods nucleate on top of pyramids, atomic steps could not explain the nucleation of all nanorods. The exact nucleation mechanisms remain under investigation.

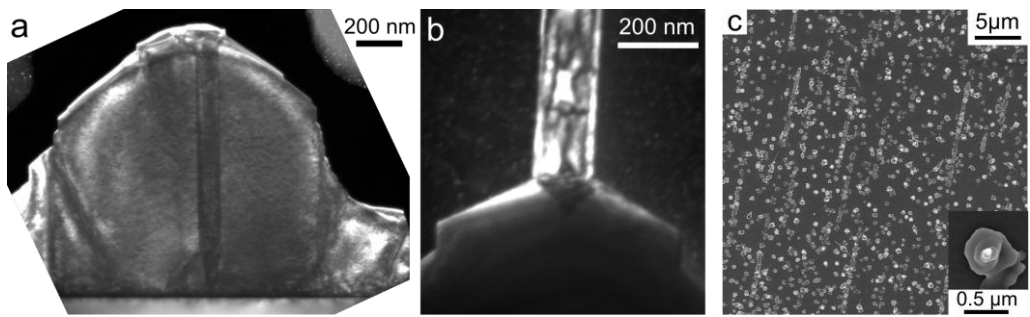

Figure 4. Nucleation of ZnO nanorods. (a) Nucleation at the sapphire / ZnO interface. The nanorod pointing out of the pyramid was removed during TEM sample preparation (b) Nucleation on top of a pyramid. (c) Top view SEM image showing alignment of pyramids and nanorods along a <11-20> direction.

3.4. Growth mechanism and defects

As it is usually observed, the orientation relationship is (Figure 5a):

$$[0001]\ Al_2O_3\ //\ [0001]\ ZnO\ \text{and}\ [-1010]\ Al_2O_3\ //\ [11\text{-}20]\ ZnO$$

leading to an in-plane mismatch of 18.4% between the sapphire substrate and ZnO. A regular network of misfit dislocations is formed at the interface in order to relax the heteroepitaxial stress (Figure 5b and c): with a thickness of 8 monoatomic layers, the wetting layer is almost totally relaxed. But no dislocations are observed in the nanorods (Figure 5d). This high crystalline quality of nanorods is confirmed by photoluminescence studies [7].

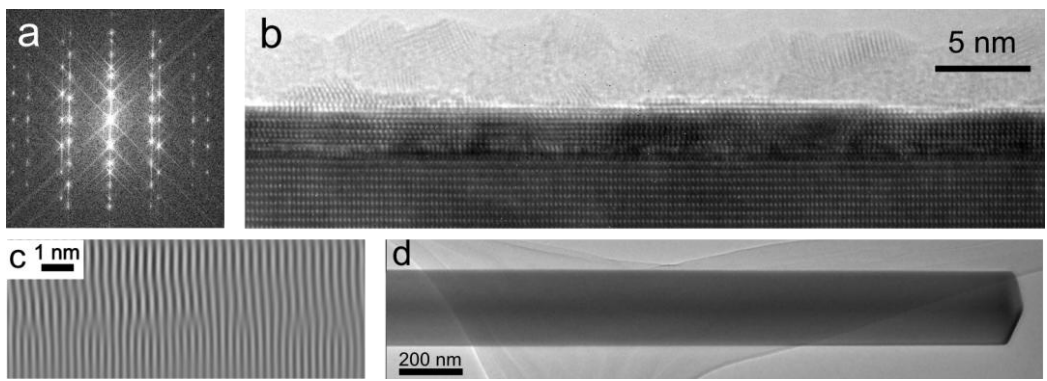

Figure 5. Stress relaxation and defects formation: (a) Fourier transform at the ZnO / sapphire interface, (b) high resolution TEM image of the ZnO wetting layer on sapphire after one min of growth, (c) [01-10] filtered image showing the network of misfit dislocations at the sapphire / ZnO interface and (d) bright field TEM image of a nanorod with no dislocations.

The nanorods nucleating at the ZnO / sapphire interface are very narrow (3 to 20 nm, Figure 6a) suggesting that few misfit dislocations are present at the base of the nanorod. Furthermore, the ratio between the nanorod length and the diameter is such that the probability for a dislocation to thread into the nanorods is low. When a nanorod has grown vertically, a pyramid nucleates around it. The pyramid grows both, laterally and vertically, and dislocations follow the growth front, threading from the nanorod base to the sides of the pyramid (Figure 6b, c and d). In the case of a pyramid with no nanorod inside, growth on the wetting layer is similar. Most dislocations in the wetting layer bend over, due to the lateral growth of the pyramid (Figure 7).

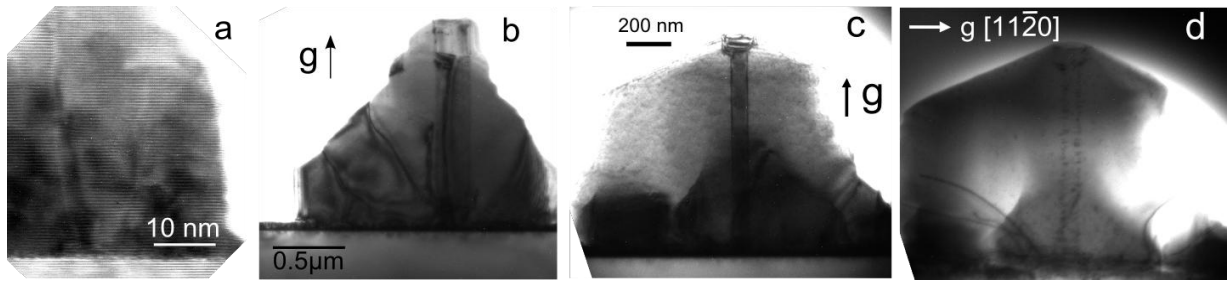

Figure 6. Nanorods surrounded by pyramids: (a) High resolution TEM image (1 min growth). (b) (c) (d) Two-beam bright field TEM images (30 min growth). The nanorods pointing out of the pyramids were removed during TEM sample preparation. Images (c) and (d) are taken on the same pyramid and with the same magnification.

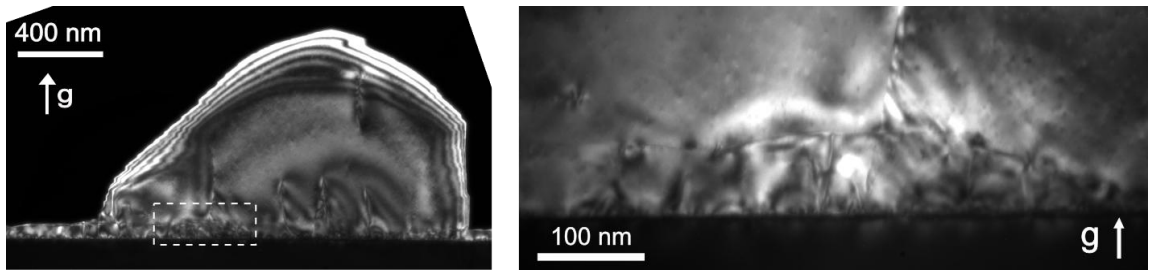

Figure 7. Defect structure of nanorods free ZnO pyramid: (0002) two beam dark field image of a pyramid on the wetting layer (left) and zoom on the interface region (right).

### 4. Conclusion
We have proposed a growth mechanism of ZnO nanorods. A 2 dimensional wetting layer covers the sapphire substrate, and nuclei appear locally to form pyramids and nanorods. Nanorods grow because some nuclei are of Zn polarity and because of the large growth rate anisotropy between [0001] and [000-1] direction. No dislocations are observed in the nanorods. Threading of dislocations from the substrate is not very probable because of the nanorod narrowness. The dislocations are not favoured energetically because of the nanorods length. Relatively few dislocations are observed inside the pyramids because of their lateral overgrowth on top of the wetting layer which bends dislocations under the pyramids.